\documentclass[traditabstract]{aa}
\usepackage{amsmath, amssymb, graphics, setspace}
\usepackage{graphicx}
\usepackage{txfonts}
\usepackage{booktabs}
\usepackage{natbib}

\renewcommand{\phi}{\varphi}

\begin{document}

\title{The orbit of 2010 TK7. Possible regions of stability for other Earth Trojan asteroids}

%%\subtitle{2010 $\text{TK}_7$}

\author{R. Dvorak\inst{1}, C. Lhotka\inst{2}, L. Zhou\inst{3} }

\authorrunning{Dvorak et al.}

\offprints{L.\ Zhou, \email zhouly@nju.edu.cn, clhotka@fundp.ac.be, dvorak@astro.univie.ac.at,
(authorlist in alphabetical order)}

\institute{Universit\"atssternwarte Wien,
 T\"urkenschanzstr. 17, A-1180 Wien, Austria,
 \and D\'epartment de Math\'ematique (naXys),
Rempart de la Vierge, 8, B-5000 Namur, Belgium,
 \and Department of Astronomy \& Key Laboratory of Modern
Astronomy and Astrophysics in Ministry of Education, Nanjing University, Nanjing 210093,
China}

\date{Received; accepted}

\abstract{Recently the first Earth Trojan has been observed (Mainzer et al.,
ApJ 731) and found to be on an interesting orbit close to the Lagrange point L4
(Connors et al., Nature 475). In the present study we therefore perform a
detailed investigation on the stability of its orbit and moreover extend the
study to give an idea of the probability to find additional Earth--Trojans. Our
results are derived using different approaches: a) we derive an analytical
mapping in the spatial elliptic restricted three--body problem to find the
phase space structure of the dynamical problem. We explore the stability of the
asteroid in the context of the phase space geometry, including the indirect
influence of the additional planets of our Solar system. b) We use precise
numerical methods to integrate the orbit forward and backward in time in
different dynamical models. Based on a set of 400 clone orbits we derive the
probability of capture and escape of the Earth Trojan asteroids 2010 TK7. c) To
this end we perform an extensive numerical investigation of the stability
region of the Earth's Lagrangian points. We present a detailed parameter study
in the regime of possible stable tadpole and horseshoe orbits of additional
Earth-Trojans, i.e. with respect to the semi-major axes and inclinations of
thousands of fictitious Trojans. All three approaches underline that the Earth
Trojan asteroid 2010 TK7 finds himself in an unstable region on the edge of a
stable zone; additional Earth-Trojan asteroids may be found in this regime of
stability.}

%%\keywords{Celestial mechanics -- Minor planets, asteroids: Earth Trojan 2010 TK7}

\titlerunning{The Earth Trojan asteroid 2010 $\text{TK}_7$}

\maketitle

\section{Introduction}

\begin{figure}[h]
\centerline{
\includegraphics[width=6.5cm,angle=0]{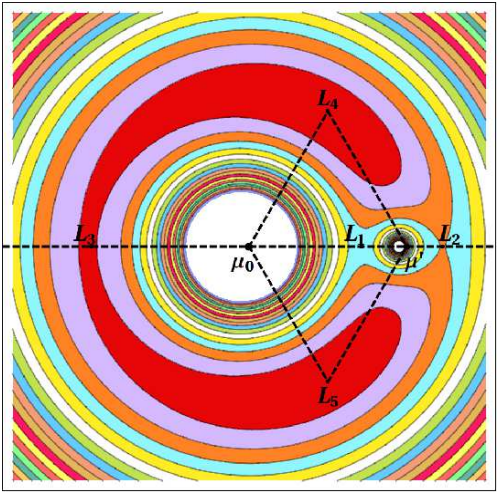}}
\caption{Geometry of the circular restricted three--body problem in
a rotating coordinate system: mass of the Sun $\mu_0$ and the Earth
$\mu'$; the unstable equilibria $L_{1,2,3}$ are located on the
x-axis; the stable points $L_4$ and $L_5$ are found $60^{\circ}$
ahead and behind the primary mass $\mu'$. The color code defines the
areas of the equipotential. In the elliptic problem the rotating
coordinate system has to replaced by a non uniformly rotating and
pulsating reference frame.} \label{schema}
\end{figure}

The giant planets Jupiter and Neptune are known to host Trojan
asteroids, and also Mars \cite{Bow90} is hosting several co-orbiting
asteroids. These kind of bodies move in the same orbit as the
planets, but around $60^{\circ}$ ahead or $60^{\circ}$ behind the
planet close to the so-called Lagrange points $L_4$ or $L_5$ (see
Fig.~\ref{schema}). It was a great surprise when quite recently the
first Earth Trojan 2010 $\text{TK}_7$ was discovered \cite{Main11}.
Although many studies have shown the principal possibility of their
existence until this event all attempts of finding one were
unsuccessful. Another small asteroid in the 1:1 mean motion
resonance (MMR) with the Earth was found earlier in 1986, by Duncan
Waldron and this asteroid (3753 Cruithne) was later identified as a
celestial body in a horseshoe orbit around both equilateral Lagrange
points of the Earth being thus not a 'real' Trojan in the sense of
its original definition (the same is true for the recently found
asteroid 2010 $\text{SO}_{16}$).

Many theoretical studies exist to establish the stability of the Lagrange
points in simplified models, i.e.  the works of \cite{Rab67, Bie84,
LhoEt2008, Erd09}, beyond many others. Ever since extensive
numerical studies have been undertaken in finding the extension of the
stability regions around the equilibrium points for the planets, e.g. in
\cite{Schw04, Dvo05, Rob05, Fre06, Dvo07}. Especially important results are
due to the  work of \cite{Mik92, Tab00, Bra02} and \cite{ Scho04}. In \cite{Mik92}
and \cite{Zha88} the authors found that Venus, Earth and Mars can host co-orbital asteroids up
to 10 Myrs. According to the investigation of \cite{Tab00} Earth's Trojans
are on stable orbits when their inclinations are relatively low ($i<16^{\circ}$);
a second stability window exists according to them for $16^{\circ}<i<24^{\circ}$.

Morais \& Morbidelli (2002) studied the orbital distribution of the
near-Earth asteroids (NEAs) that experience the episode of being in
the 1:1 MMR with the Earth. In a most recent study \cite{Schw11} the
possibility of captures of asteroids by the terrestrial planets into
the 1:1 MMR was investigated and many temporary captures including
jumping Trojans \cite{Tsi00} were found; as we will see also 2010
$\text{TK}_7$ has such an interesting captured orbit.

In our investigations -- initiated by the finding of 2010
$\text{TK}_7$ -- we concentrate on Earth Trojans. Three different
approaches should clarify the stability problem of these asteroids:
in Chapter 2 we perform analytical studies in a simplified model,
valid on short time scales. We investigate the phase phase structure
and the influence of additional perturbations in the simplified
model. In Chapter 3 we study in great detail the actual orbit of
2010 $\text{TK}_7$ together with 400 clone orbits to obtain a better
statistics to state the probability of capture and escape of this
Trojan asteroid. To this end we perform in Chapter 4 an extensive
numerical investigation of the parameter space relevant to determine
the stable and unstable regions close to the equilibrium point
$L_4$~\footnote{both Lagrange points have in principal the same
dynamical behaviour as we know from earlier studies e.g.
\cite{Nes09, Zho09}} and show the interplay between secular
resonances and the stability of motion. The summary and conclusions
of the present study are found in Section 5.

\section{A symplectic mapping model}

\begin{figure}
\centering
\includegraphics[width=8.5cm, angle=0]{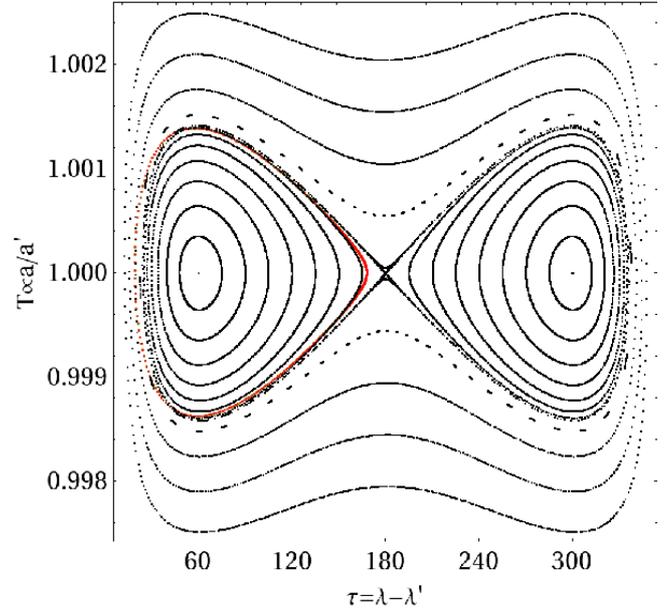}
\caption{Phase portrait - projected to the \((\tau ,T)\)-plane. The
Earth is located at \((0,1)\), the fixed points \(L_4\), \(L_3\),
\(L_5\) (corresponding to the equilibria of the averaged system) are
situated at \((60^\circ,1)\), \((180^\circ ,1)\) and
\((300^\circ,1)\), respectively. The projection of the mean orbit of
the asteroid 2010 \(\text{TK}_7\) for \(\pm 6000\) years is shown in
red.} \label{phapor}
\end{figure}

The Hamiltonian describing the motion of 2010 \(\text{TK}_7\) in
the spatial elliptic restricted three--body problem
(Sun-Earth-asteroid) takes the form:

\begin{equation}
\label{Ham1}
H=H_{\text{Kep}}+T+\mu 'R(a,e,i,\omega ,\Omega ,M,M';p')
\end{equation}

\noindent where \(H_{Kep}\) denotes the Keplerian part, the function \(\mu 'R\)
is the perturbation due to the Earth with mass \(\mu '\) and the variable \(T\)
is the action conjugated to time (assuming, that the mean motion \(n'\) of the
Earth is equal to one). Moreover, \(a,e,i,\omega ,\Omega ,M\) are the
semi-major axis, the eccentricity, the inclination, the perihelion, the
longitude of the ascending node and the mean anomaly of the asteroid,
respectively, while \(M'\) denotes the mean anomaly of the Earth.

In contrast to the classical expansion techniques of the
perturbing function \(R\) we do not replace the remaining orbital
parameters of the Earth \(p'=(a',e',i',\omega ',\Omega ')\) with
their numerical values but rather keep them as free parameters in
the ongoing calculations. The simple reason for that is to be able
to investigate their influence on the dynamics of the massless
asteroid with time, i.e. to see their influence on the phase space
geometry, later on. For the mapping approach we set the system of
constants (gravitational constant \(G\) and total mass of the
system) equal unity such that \(G(\mu_0+\mu')=1\) which also implies
that \(a'\) as well as \(a\) are close to unity and in these
dimensionless units one revolution period of the Earth takes
\(P=2\pi\). Furthermore, we use as action-angle variables the
modified Delaunay variables which are defined in terms of the
classical Delaunay variables by:
 \begin{eqnarray}
&&l=M, \ g=\omega , \ h=\Omega \nonumber \\
&&L=\sqrt{a}, \ G=L\sqrt{1-e^2}, \ H=G \cos (i) \nonumber
 \end{eqnarray}
\noindent as
 \begin{eqnarray}
 &&\lambda _1=l+g+h, \ \lambda _2=-g-h, \ \lambda _3=-h \nonumber \\
&&\Lambda _1=L, \ \Lambda _2=L-G, \ \Lambda _3=G-H \nonumber
 \end{eqnarray}
\noindent (and similar for the primed variables of the Earth). In
this setting, the Hamiltonian (\ref{Ham1}) becomes:
 \begin{equation}
 \label{Ham2}
 H=-\frac{1}{2\Lambda _2{}^2}+T+\mu 'R\left(\lambda ,\Lambda ,\lambda _1';q'\right)
 \end{equation}
\noindent where we used the short-hand notation \(\lambda =\left(\lambda _1,\lambda _2,\lambda _3\right)\) and
 \(\Lambda =\left(\Lambda _1,\Lambda _2,\Lambda _3\right)\). In addition we abbreviate the parameter vector
\(q'=\left(\lambda _2',\lambda _3',\Lambda _1',\Lambda _2',\Lambda
_3'\right)\) to define the vector of modified Delaunay variables for
the Earth. Next, we implement a symplectic change of coordinates and
momenta suitable to describe the motion of the asteroid close to the
\(1:1\) MMR. We define the resonant angle and conjugated momenta as:
 \begin{equation}
\label{resarg}
\tau =\lambda _1-\lambda _1', \ T=\Lambda _1
 \end{equation}
\noindent while the other variables transform by the identity:
 \begin{equation}
\lambda _2=\phi , \ \lambda _3=\theta , \ \Lambda _2=\Phi , \ \Lambda _3=\Theta \nonumber
 \end{equation}
\noindent  and \(T'=\Lambda _1-\Lambda _1'\). We replace \(\lambda
_1\) in (\ref{Ham2}) according to (\ref{resarg}) and implement the
standard averaging procedure over the fast angle \(\lambda '\) via
the formula:
 \begin{equation}
\label{Ham3}
\tilde{H}=-\frac{1}{2T^2}+
\frac{1}{P}\int _0^{P}\mu 'R\left(\lambda ,\Lambda ,\lambda _1';q'\right)\mathrm{d}\lambda' \ ,
 \end{equation}
 \noindent with $P=2\pi$, to get the averaged Hamiltonian function of the form \(\tilde{H}=\tilde{H}(\psi ,\Psi ;q')\) with
\(\psi =(\tau ,\phi ,\theta )\) and \(\Psi =(T,\Phi,\Theta )\). Our
aim is to construct a symplectic mapping which transforms the state
vector \(\left(\psi _k,\Psi _k\right)\equiv \left(\tau
_k,\phi_k,\theta _k,T_k, \Phi _k,\Theta _k\right)\) at discrete
times \(k\) (multiples of \(P\)) to the state vector \(\left(\psi
_{k+1}, \Psi _{k+1}\right)\) at times \(k+1\). For this reason we
define the generating function
 \begin{eqnarray}
W_{q'}=W_{q'}\left(\tau _k,\phi _k,\theta _k,T_{k+1},\Phi _{k+1},\Theta _{k+1};q'\right)=\nonumber \\
\psi _k\cdot J_{k+1}+2\pi \tilde{H}\left(\psi _k,J_{k+1};q'\right) \ , \nonumber
 \end{eqnarray}
 \noindent where the symbol $\cdot$ is the dot product. As it has
been shown and already used in \cite{Had92, Had99, Had2000} or
\cite{Fer1997} the generating function defines a mapping of the
form:
 \begin{eqnarray}
\label{map}
&&\tau _{k+1}=\frac{\partial W_{q'}}{\partial T_{k+1}}, \ \
\phi _{k+1}=\frac{\partial W_{q'}}{\partial \Phi _{k+1}}, \ \
\theta _{k+1}=\frac{\partial W_{q'}}{\partial \Theta _{k+1}}, \nonumber \\
&&T_k=\frac{\partial W_{q'}}{\partial \tau _k}, \ \
\Phi _k=\frac{\partial W_{q'}}{\partial \phi _k}, \ \
\Theta _k=\frac{\partial W_{q'}}{\partial \theta_k} \ \ .
 \end{eqnarray}

System (\ref{map}) defines a symplectic change of coordinates on the
Poincar\'{e} surface of section obtained by the averaged system
defined by (\ref{Ham3}). The set of variables \((\Psi ,\psi )\) is
related to the mean orbital elements of the asteroid 2010
\(\text{TK}_7\) via another generating function (not derived here)
which defines the averaging process given in (\ref{Ham3}). The
system (\ref{map}) therefore describes the evolution of the mean
orbital elements of the asteroid at discrete times \(t=k\cdot P\).
Note, that the system is implicit to preserve the Hamiltonian
structure of the original problem. For given initial conditions
\(\left(\psi _{k=0},\Psi _{k=0}\right)\) it can either be iterated
by solving the system of difference equations implicitly for
\(\left(\Psi _{k+1},\psi _{k+1}\right)\) or by more sophisticated
procedures as described e.g. in \cite{Lho09}.

A typical projection of the phase portrait to the $(\tau,T)$-plane
is shown in Fig.~\ref{phapor}. The lines in black were obtained by
varying $\tau$ within $(0^\circ,360^\circ)$ along $T=1$. The plot
also shows the mean orbit of 2010 TK$_7$ in red. It has to be
compared with Figure 2 of \cite{Con11}: while the authors derive the
red curve by numerical averaging, the averaged orbit of the present
approach is based on Eq.~(\ref{map}).

As it is well-known, e.g. \cite{MorMor2002, Bra02, Scho05}, the
influence of the other Solar system bodies, i.e. the direct
influence of the major planets, affects the motion of the asteroids
on long time scales. The mapping does not take into account these
direct effects. Thus, the mean orbit as shown in Fig.~\ref{phapor}
can only been seen as a first approximative solution of the problem.
Moreover, the simulations were done using low order expansions of
the perturbing function, since we wanted to keep the dependency of
the model on the orbital parameters of the Earth. The resulting
mapping model is therefore only valid within a good convergence
regime of the Fourier-Taylor series expansions used to approximate
\(R\) in (\ref{Ham1}). In addition, since the orbital parameters of
the asteroid 2010 \(\text{TK}_7\) may reach high values (as found
from numerical simulations) the error of the approximation of the
series expansions may exceed the cumulative effect of the
perturbation effects due to the Solar system and may also change the
picture dramatically on longer time scales. An estimate of the error
terms and higher order series expansions together with their
influence on the long term dynamics of the asteroid 2010
\(\text{TK}_7\) is currently under investigation.

\section{The orbit of 2010 $\text{TK}_7$}

The Earth Trojan 2010 $\text{TK}_7$ fundamentally is a NEA, and its
orbital elements (listed in Table~\ref{tab-orb}) can be found also
on the AstDyS ({\it Asteroids - Dynamic Site})
website~\footnote{http://hamilton.dm.unipi.it}. At first glance,
this object has a large eccentricity ($\sim 0.19$) and it is
reasonable to suspect that it's on a unstable orbit. To check the
orbital stability of this object, we perform numerical simulations
of the orbit.

We adopt two dynamical models in our simulations. Both models
contain the Sun and 8 planets from Mercury to Neptune. They differ
from each other by different settings of the Earth-Moon system. In
one model, the Earth-Moon system is simply treated as one mass point
with the combined mass of the Earth and the Moon at the barycenter
of the system. And in the other model, the Earth and the Moon are
regarded as two separated objects as in reality. Hereafter we denote
the former by EMB model and the latter by E+M model for short. By
comparing these two models, we may testify the reliability of the
model adopted later in this paper and also by \cite{Con11} in their
numerical simulations.

\begin{figure}[h]
\centerline{
\includegraphics[width=8.5cm,angle=0]{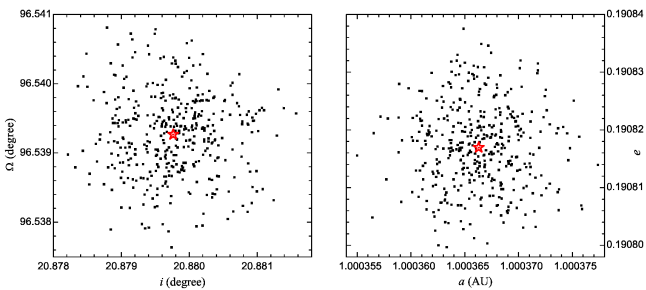}}
\caption{The initial conditions of clone orbits. Dots are for 400
clones, and the nominal orbit is indicated by the red star. We show
here in two panels only the inclination, ascending node, semimajor
axis and eccentricity.} \label{InitCond}
\end{figure}

Considering the uncertainties in observation and orbital
determination, it is necessary to study clone orbits within the
error bars. Beside the nominal orbit, 400 clone orbits are generated
using the covariance matrix given by the AstDyS website. As listed
in Table~\ref{tab-orb}, the errors are very small, therefore the
initial conditions of the clone orbits are very close to the nominal
orbit as shown in Fig.~\ref{InitCond}. We show here the distribution
of $a, e, i$ and $\Omega$, the other two elements $M, \omega$ not
illustrated here have similar distributions according to the
corresponding uncertainties listed in Table~\ref{tab-orb}.

\begin{table}
\caption{The Keplerian elements of 2010 $\text{TK}_7$ given at epoch
JD2455800.5. The elements and $1-\sigma$ variations are from the
AstDyS (see text).
The covariance matrix is also taken from the same website. }
 \begin{tabular}{|c|c|c|c|}
 \hline
  & Value & 1-$\sigma$ variation & Unit \\
 \hline
  $a$ & 1.00037 & $2.546\times 10^{-7}$ & AU \\
  $e$ & 0.190818 & $9.057\times 10^{-7}$ & \\
  $i$ & 20.88  & $7.274\times 10^{-5}$ & deg. \\
  $\Omega$ & 96.539 & $1.842\times 10^{-4}$ & deg. \\
  $\omega$ & 45.846 & $2.309\times 10^{-4}$ & deg. \\
  $M$ & 217.329 & $1.848\times 10^{-4}$ & deg. \\
 \hline
 \end{tabular}
 \label{tab-orb}
\end{table}

Starting from the initial conditions generated in the above way, we
integrate their orbits up to 1 million years (Myr) in both
directions (forward to the future and backward to the past). We used
the integrator package {\it Mercury6} \cite{cham99} and verified
some results using the Lie-integrator \cite{Han84}. The comparison
between the results from different integrators shows that the
results are consistent with each other.

For an asteroid on a Trojan-like orbit, the critical angle
(resonant angle) is the difference between the mean longitudes of
the asteroid $\lambda$ and the corresponding host planet (here
it's the Earth-Moon system) $\lambda^\prime$, like the $\tau$ in
Eq.~(\ref{resarg}). In the EMB model the resonant angle is
$\lambda - \lambda_{\rm EMB}$, while in the E+M model we calculate
the position and velocity of the barycenter of the Earth-Moon
system from the orbital elements of the Earth and the Moon, and
then compute the resonant angle in the same way as in the EMB
model. When the resonant angle $\lambda - \lambda^\prime  =
\lambda - \lambda_{\rm EMB}$ librates around $60^\circ$, i.e.
$0^\circ < \lambda - \lambda^\prime < 180^\circ$, the asteroid is
said to be an $L_4$ Trojan, when it librates around $-60^\circ$ or
$300^\circ$, i.e. $180^\circ < \lambda - \lambda^\prime <
360^\circ$ it is an $L_5$ Trojan, when the asteroid librates with
an amplitude larger than $180^\circ$ it is on a horseshoe orbit,
and finally when $\lambda - \lambda^\prime$ circulates the
asteroid leaves the Trojan-like orbit.

\begin{figure}[h]
\centerline{
\includegraphics[width=8.cm,angle=0]{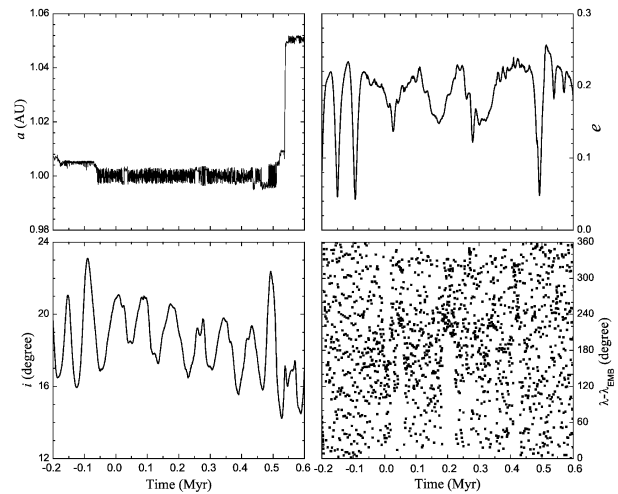}}
\caption{The temporal evolution of the nominal orbit in 1 million
years, both forward and backward. In four panels we show the
evolutions of the semimajor axis (upper left), eccentricity (upper
right), inclination (lower left) and the resonant angle (lower
right). }
 \label{NomOrbEvo}
\end{figure}

\begin{figure}[h]
\centerline{
\includegraphics[width=7.5cm,angle=0]{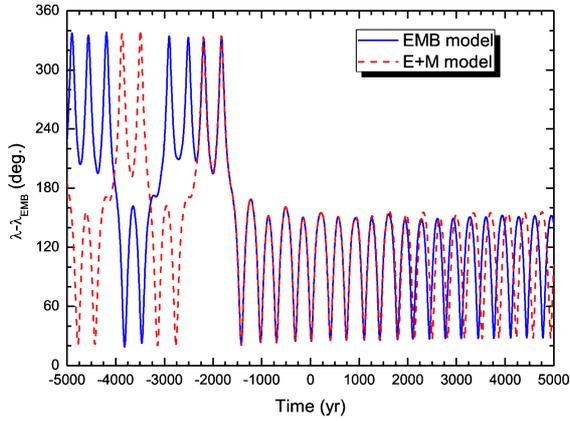}}
\caption{The resonant angle of the nominal orbit in two models. The
difference between them is distinguishable very soon ($\sim 2000$
years) mainly due to the chaotic character of the motion (see
discussion in text).} \label{NomOrb2Mod}
\end{figure}

In Fig.~\ref{NomOrbEvo}, we illustrate the temporal evolution of
the nominal orbit in the E+M model. The first impression obtained
from its behaviour may be that it's a temporal Earth Trojan, i.e.
judging from the resonant angle it was not an Earth Trojan $0.055$
Myr before and it will not stay on a Trojan-like orbit after
$0.37$ Myr. The period of its being an $L_4$ Trojan is even much
shorter, less than 2,000 years in the past and less than 17,000
years in the future (as partly shown in Fig.~\ref{NomOrb2Mod}).
During the period of being a Trojan, the semimajor axis $a$ shows
regular variations librating around 1.0 AU with a small amplitude,
however, the variations of eccentricity $e$ and inclination $i$
reveal the chaotic character of the motion, as shown in
Fig.~\ref{NomOrbEvo}.

The chaotic character of the motion implies not only a sensitive
dependence on the initial conditions but also a sensitive
dependence on the dynamical model. To compare the motions in two
models, i.e. the EMB model and E+M model, we show as an example in
Fig.~\ref{NomOrb2Mod} the evolution of the resonant angle of the
nominal orbit in both models. Around the starting point, two
curves representing the motions in two models are almost the same
so that they overlap each other exactly. But the difference
between them becomes distinguishable only after about 2,000 years
in both directions. This difference on one hand arises from the
different settings of the models, on the other hand, it is due to
the chaotic character of the motion. In this sense, it is
impossible to draw any convincible conclusion about these two
dynamical models on long time scales by comparing just the motions
of an individual orbit in two models. So we turn to analyze
statistically the four hundred clone orbits below.

As we mentioned above, 400 clone orbits are calculated in both
models and in both directions of time. At starting moment $t=0$
these clones are around the $L_4$ Lagrange point as the asteroid
2010 $\text{TK}_7$, but in the evolution, the objects may jump
from the $L_4$ region to $L_5$ region, or they may move from
tadpole orbit to horseshoe orbit, and even, they may escape from
the Trojan-like orbit (the 1:1 MMR). To examine the motion, we
check the resonant angle $\lambda - \lambda_{\rm EMB}$ at each
step during our simulations. When $\lambda - \lambda_{\rm EMB}$
for the first time is larger than $180^\circ$, the object is
regarded as leaving the $L_4$ region, and we denote the time at
this moment as $t_1$. When $\lambda - \lambda_{\rm EMB}$ reaches
for the first time $360^\circ$, the object escapes the 1:1 MMR,
and we denote this moment as $t_2$.

\begin{figure}[h]
\centerline{
\includegraphics[width=7.5cm,angle=0]{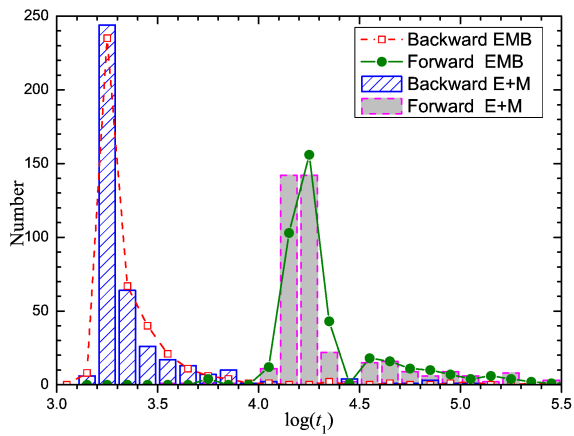}}
\caption{The distribution of the time when clone orbits leave from
current $L_4$ region. The escape time from the tadpole region
is given in years in logarithm scale. }
 \label{TimeLeavL4}
\end{figure}

\begin{figure}[h]
\centerline{
\includegraphics[width=7.5cm,angle=0]{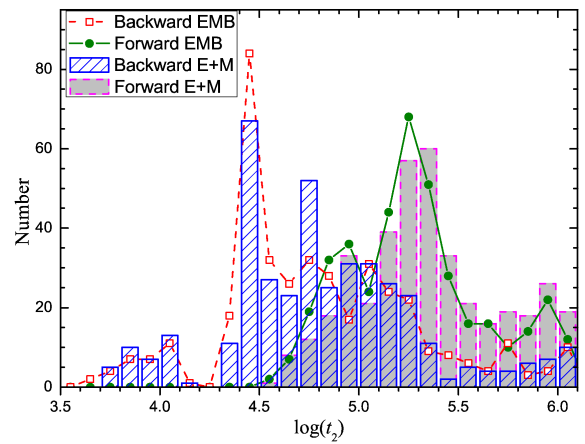}}
\caption{The distribution of the time when clone orbits escape from
the 1:1 MMR, both in forward and backward integrations. The escape
time is given in years in logarithm scale.}
 \label{TimeEscTro}
\end{figure}

The $t_1$ and $t_2$ of 400 clone orbits are summarized in
Fig.~\ref{TimeLeavL4} and Fig.~\ref{TimeEscTro}. For $t_1$, most
of the clones will leave the $L_4$ region in about $1.5 \times
10^4$~years in future, while the backward integrations indicate
that the clones ``entered'' the $L_4$ region in about $1.8 \times
10^3$~years ago (Fig.~\ref{TimeLeavL4}).

In the backward integrations, the earliest escape from the $L_4$
region happens at $t\sim 1440$~years both in the E+M and EMB models.
But there are only a few such orbits. Most of them escape in the
time range between $t\sim 1680$ and $t \sim 1820$~years which causes
the peak centered at $\log(t_1)=3.25$ in Fig.~\ref{TimeLeavL4}. In
fact, from Fig.~\ref{NomOrb2Mod}, we may derive the libration period
of the resonant angle is around 350 years\footnote{Due to the large
libration amplitude, this value is nearly 50 percent larger than the
synodic period of a tadpole orbit which can be estimated using the
formula $(27\mu/4)^{-1/2}$ \cite{mur99}. For an Earth Trojan, the
mass ratio $\mu=3.04\times 10^{-6}$ and the period is $\sim
220$~years.}, and we also note that the shift of the orbit from
$L_4$ to $L_5$ region happens when the resonant angle reaches the
maximum in a libration period. The libration period and amplitude
must be tuned by other periodic effects (e.g. secular resonances).
After a number of complete periods of evolution, the resonant angles
of some clone orbits reach their maxima, and they do not librate
back but escape from there towards the neighbourhood of $L_5$. After
another complete libration period, many more clone orbits escape in
the same way as before. That's why the escapes from the $L_4$ region
seem to happen more or less suddenly (at $t\sim 1440+350=1790\approx
10^{3.25}$~years). All clone orbits are in a tiny region confined by
the error bars, so that they suffer nearly the same dynamical
effects in a short timespan. For those clones staying in the region
longer, they spread in the phase space and thus suffer different
dynamical effects and consequently their escape times diverge.

The distributions of $t_1$ for the EMB model and the E+M model in
Fig.~\ref{TimeLeavL4} match each other very well, therefore we may
draw another conclusion here that the difference between these two
dynamical models is ignorable, or in other words, the simplified EMB
model is a reasonable and reliable model for investigating the Earth
Trojans' long term dynamics. Particularly, we would like to note
that the asteroid 2010 $\text{TK}_7$ and the clones all have very
large libration amplitudes around the $L_4$ and/or $L_5$ point (see
for example Fig.~\ref{NomOrb2Mod}), i.e. they may approach to the
Earth-Moon system (mean longitude difference between them may be
less than $20^\circ$). But in fact, even when an Earth Trojan
approaches the Earth to within $10^\circ$ apart in mean longitude,
the distance between the asteroid and the Earth is about 0.18~AU,
which is still about 70 times larger than the distance between the
Earth and the Moon. Taking into account the inclination of the Earth
Trojan's orbit, the distance between the Trojan and the Earth is
even larger.

As for $t_2$ distribution in Fig.~\ref{TimeEscTro}, we see the peak
of escaping time in the backward integrations locates at $\sim 3.0
\times 10^4$~years. As we mentioned above, the Earth Trojan 2010
$\text{TK}_7$ should be a temporal Trojan. The capture of this
asteroid onto the 1:1 MMR with the Earth happened most probably
30,000 years ago. In the forward integrations, most of clones will
escape from the MMR around $2.0 \times 10^5$~years after. The total
lifetime of this asteroid in the 1:1 MMR is less than 0.25 Myr. And
again, the two models E+M and EMB give nearly the same
distributions.

In our calculations, the episode of clones being $L_4$ Earth Trojans
lasts typically a little less than $\sim 17,000$ years. As for the
lifetime of clones staying in the 1:1 MMR, there are 10 clones
surviving 1 Myr in the backward integrations in both models, while
for the forward integrations, 12 in the EMB model and 19 in the E+M
model stay in the MMR till the end of integrations. But none of the
clones survives in both temporal directions. In one word, less than
5 percent of clones stay in the MMR up to 1 Myr. Morais \&
Morbidelli (2002) calculated the probability of a NEA being captured
into the 1:1 MMR (``coorbital orbit'' in their paper), and they
found that each episode of a NEA being coorbital on average is
25,000 years and none lasts longer than 1 Myr. Our result does not
conflict with their conclusion, because our calculations are for the
individual asteroid 2010 TK$_7$, and its eccentricity ($\sim 0.2$)
is smaller than the typical eccentricity in their samples (most of
them have $e>0.28$).

Some fluctuations in the distribution of $t_1$ and $t_2$ (by
``fluctuation'' we mean more peaks deviated from a trivial normal
distribution) can be found in both models in Fig.~\ref{TimeLeavL4}
and Fig.~\ref{TimeEscTro}. In these figures, the time is given in
logarithm to include a wide time range. If we plot the time
linearly, these fluctuations show some periodic character, implying
that some periodic mechanisms (e.g. secular resonances) are
affecting the libration amplitudes. However to depict a secular
resonance map is a long story, we would like leave it to a separate
paper but show first results in Section 4.3.

\section{Determination of the stable regions}

As was pointed out in the previous sections the orbit of 2010 $\text{TK}_7$
is not very stable. The goal of this part of our investigation is to determine
the largeness of the stable regions around the stable equilibrium points. This
can be done in a realistic dynamical model only with the aid of numerical
integrations where many fictitious Trojans are checked concerning the stability
of their orbits. The dynamical model used was a 'truncated' planetary system
with the planets Venus to Saturn (Ve2Sa). The simple reason for not taking
into account all the planets is that it would at least require four times
longer CPU times on our computers. It is mainly due to the orbit of the
innermost planet Mercury which demands for a step size of one quarter of
the one of Venus\footnote{The inclusion of Uranus and Neptune would not
dramatically change this integration time}. Test computation for selected
stable and unstable orbits showed that the qualitative behavour of such orbits
is not a different one and consequently also the largeness of the stable
region can be regarded as the 'real' one. It means that the gravitational
perturbation of Mercury on a Trojan for this study is neglectable. The Moon
was not explicitly integrated but the barycenter Earth Moon with the respective
masses was taken as one body. As has been shown in Chapter 3 there are only
very small differences when one integrates the Trojan orbits in these models,
on long time scales. On the contrary Jupiter's direct perturbation on an Earth
Trojan is very large and cannot be neglected especially for large librations
which bring the Trojan close to the Lagrange point $L_3$.

The integration method was the Lie-series method which we used
already quite often in similar investigations (e.g. for Neptune
\cite{Zho09, Zho11} and recently also Uranus' Trojans
\cite{Dvo10}). This method is based on a work by \cite{Han84,
Del84} and  \cite{Lic84} and was slightly modified for our problem
(see also \cite{Egg10}). It has an automatic step-size control and
turned out to be fast and precise.

The initial conditions were chosen in the following way: for the fictitious
Trojans the orbital elements $M$, $\Omega$, $e$ were set to the one of the Earth.
The perihelion was always $\omega_{Earth}+ 60^{\circ}$ \footnote{which means that
the position was the Lagrange point $L_4$}, the semimajor axes for the Trojan
was set to slightly smaller and larger values to cover the stable region (along
the connecting line between Sun and the Lagrange point). We started with
integrations in the plane of the orbit of the Earth and changed in additional
runs the inclinations up to $i=60^{\circ}$.

As to distinguish between stable and unstable orbits we used
different indicators: the most straightforward was to check the
eccentricity of the object because there turned out to be a sharp
cut for $e>0.3$: any orbit which achieved this value during the
integration left the area around the Lagrange point (checked by
its distance to this point). In another test we computed directly
the escape times and finally we also computed the libration width
of the Trojan which is a well established check of stability for a
Trojan. The length of integration was thoroughly chosen; although
we had a computer grid available the main computations covered
only $10^7$ years but some tests has been undertaken up to $10^8$
years.

\subsection{The stable regions for different time scales}

\begin{figure}[h]
\centerline{
\includegraphics[width=8cm,angle=270]{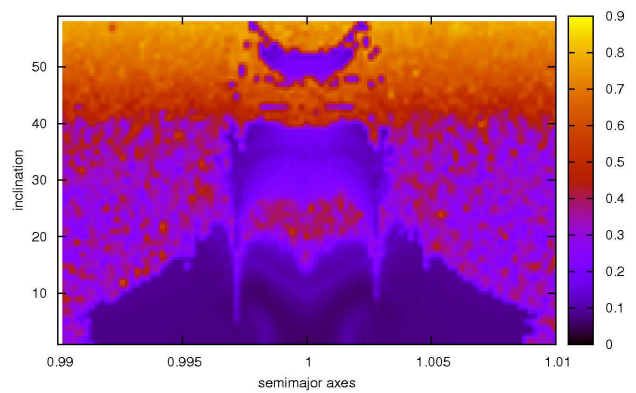}}
\caption{Stability diagram using $e_{max}$ for the window between $i <
  60$ (y-axes) for different
  initial semimajor axes (x-axes) of the fictitious Trojans; the color stands
  for the maximum eccentricity during the integration of $10^6$ years.}
\label{all-60}
\end{figure}

The stability diagram for a cut through the Lagrange point $L_4$ is
shown in Fig.~\ref{all-60} where we plotted the maximum eccentricity of the
orbits of 100 massless fictitious Trojans for different values
of the semimajor axes (x-axes) versus the inclinations of the Trojan
(y-axes). One can see that for small inclinations the stable region extends
in semimajor axes between $ 0.99 \leq a \leq 1.01$, then there is
a decrease of the largeness of the region visible up to about $i=20^{\circ}$. Inside this
region two finger-like slightly less stable regions are visible on both sides
of the a V-shaped structure centered at $a=1$ leading from  $15^{\circ} \leq i \leq
20^{\circ}$. Then an unstable strip follows to about  $i=26^{\circ}$ with some
connection to a larger rectangular stable region with $0.997 \leq a \leq 1.003$ and
 $26^{\circ} \leq i \leq 40^{\circ}$. Stable symmetric fingers inside this
region are clearly visible and build a continuation of the unstable fingers
around $a\simeq0.997$ and $a\simeq1.003$ for the large stable area for $i<20^{\circ}$.
For larger inclinations on   $i > 40^{\circ}$ the unstable regions extends for all
values of the semimajor axes with the exception of a small U-shaped stable window
around $i = 50^{\circ}$. Note that inside this unstable region (red to yellow)
sometimes small stable island seem to appear on both sides of the stable region
$i < 40^{\circ}$ which - after a longer integration - disappear.

Concerning the structures visible in the figure mentioned above we
note that similar ones have been discovered for the Trojan regions
of the outer planets in the papers by \cite{Mit02} and \cite{Nes02b}
for larger eccentricities. In our paper we investigated the
stability for larger inclinations; the different features inside
stable regions for Earth Trojans are comparable to the ones found by
\cite{Zho09} for the Neptune Trojan region. The instabilities inside
the stable region are caused by secular resonances -- as it has been
shown in detail for the Jupiter Trojans by \cite{Rob06}. In our
article we just want to show -- for the moment -- the complicated
structures. To make a detailed analysis concerning the resonances
like in the paper by \cite{Zho09} or even the former mentioned
article by \cite{Rob06} takes quite a long time; we started this
work already.

\begin{figure}[h]
\centerline{
\includegraphics[width=8cm,angle=270]{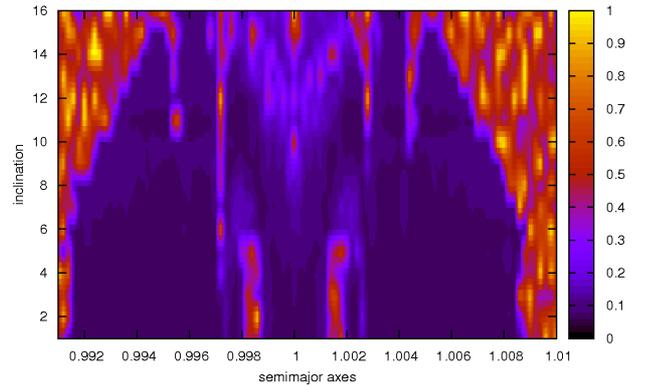}}
\caption{Caption like for Fig.~\ref{window}.}
\label{i1-16}
\end{figure}

In a next step we extended the integration to $10^7$ years in the same dynamical
model Ve2Sa. The region for small inclinations is still stable with the same
extension in semimajor axes (Fig.~\ref{i1-16}). New
features, unstable vertical strips, appear for different values of $a$ and increasing
values of the inclination. For $a\simeq0.997$ and $a\simeq1.003$ these regions were
already visible as being less stable in
the former plot (Fig.~\ref{all-60}) for small inclinations. New unstable vertical
fingers appear also for $a \sim 0.995$ and $a \sim 1.005$ and  $10^{\circ} < i <
16^{\circ}$. A new characteristic is the appearance of unstable fingers for  $a \sim 0.998$ and
$a \sim 1.002$ with  $i < 6^{\circ}$. The V-shaped unstable region centered at
$a=1$ was already visible in Fig.~\ref{all-60} but the detailed structures become
more clear using a longer integration time.

In the next Fig.\ref{window} the stable window for  $28^{\circ} <
i < 40^{\circ}$ was studied separately (also for $10^7$ years).
Again one can see that the most stable regions visible through the
dark blue -- indicating small e-max values -- are close to the
edges in $a$. These edges are quite irregular and some smaller
unstable islands appear inside; the whole stable region is
somewhat tattered. What we do not show here is the disappearance
of the U-shaped island for $i\sim50^{\circ}$ for a longer
integration time.

We note that we disagree, in some aspects, with the results given
by \cite{Tab00} who claim that stable regions for Earth Trojans
are possible for $24^{\circ} < i < 34^{\circ}$; the stable window
we found is shifted outwards to larger inclinations. We also
disagree with the later paper by \cite{Bra02} where they say that
orbits with an inclination of  $12^{\circ}  < i < 25^{\circ}$ on
are unstable. In the last work the authors quite nicely determine
the secular frequencies involved leading to unstable motion. We
already started additional computations where the determination of
the resonances are undertaken, but their detailed analysis will be
shown in a longer article (in preparation).

\begin{figure}[h]
\centerline{
\includegraphics[width=8cm,angle=270]{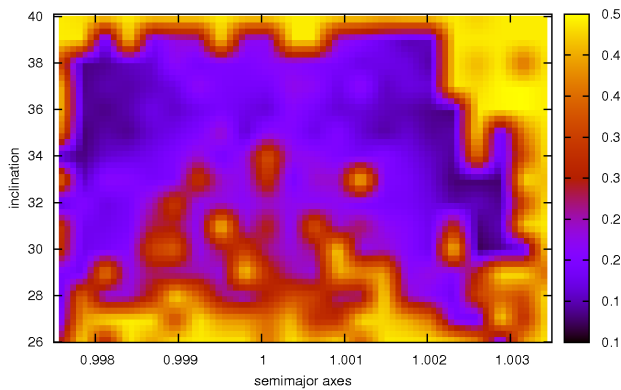}}
\caption{Stability diagram using $e_{max}$ for the window between $26 < i <
  40$ (y-axes) for different
  initial semimajor axes of the fictitious Trojan; the color stands
  for the maximum eccentricity during the integration of $10^7$ years.}
\label{window}
\end{figure}

\subsection{The libration amplitudes}

To determine the libration width of the region we choose for the semimajor
axes a grid of $0.995 \leq a \leq 1.005$ for 50 different fictitious bodies equally
distributed in the mentioned interval where the inclination was set to values
$0^{\circ} \leq i \leq 56^{\circ}$. The integration time was only $10^6$ years for
this study. The colors (from blue to yellow)  indicate the
amplitude of libration; we can see a well defined stable region in the range
$0.997 \leq a \leq 1.003$ for inclinations $i \leq 19^{\circ}$. This rectangular like
region seems to contradict the results shown in Fig.~\ref{all-60}, but a
closer look shows -- through the color of dark yellow -- that the libration angle is
in the order of $170^{\circ}$; thus on the edge we have horseshoe
orbits which enclose both equilateral Lagrange points. An unstable strip then
arises $20^{\circ} \leq i \leq 24^{\circ}$ where close to $a \sim 1.0005$ and $ i \sim
21^{\circ}$ the 2010 $\text{TK}_7$ is located. Then we see again a large almost
rectangle like stable region $25^{\circ} \leq i \leq 39^{\circ}$ for the same range
of $a$ as the first stable region of tadpole orbits for only slightly
inclined Trojan orbits. Note that the edges are sharp and no horseshoe orbits
are located there.
From  $i = 40^{\circ}$ no more stable orbits exist with the exception of a
small stable window with tadpole orbits for $i \sim 50^{\circ}$; these orbits
will be unstable for longer integration time as indicated above.

\begin{figure}[h]
\centerline{
\includegraphics[width=8cm,angle=270]{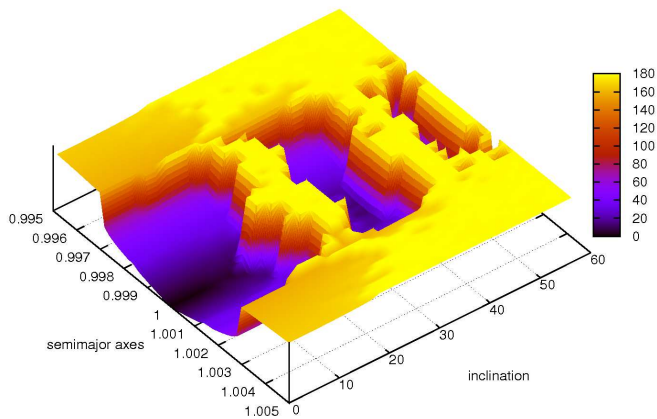}}
\caption{Libration amplitudes --  defined as half of the total
libration angle --  in the region close to $L_4$; the color stands
  for the largeness of the libration.}
\label{lib}
\end{figure}

\begin{figure}[h]
\centerline{
\includegraphics[width=6.5cm,angle=270]{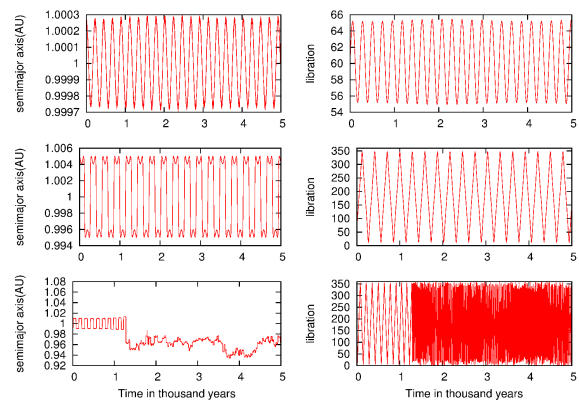}}
\caption{The temporal evolution of three orbits: a tadpole (upper graphs) a
  horseshoe (middle graphs) and an escaping orbit (lower graphs)
In six panels we show the
evolutions of the semimajor axis (left graphs) and the libration angles (right
graphs) during 5000 years. }
\label{3orbs}
\end{figure}

To show the different kinds of orbits we depict three different ones
for an inclination of the Trojan's orbit $i=1^{\circ}$: a tadpole
orbit deep inside the stable region (Fig.~\ref{3orbs} upper graphs),
a horseshoe orbit close to the border of stability (Fig.~\ref{3orbs}
middle graphs) and an escaping orbit outside the stable region after
initially being in a horseshoe orbit (Fig.~\ref{3orbs} lower
graphs). On the left pictures of this figure the semimajor axes show
a periodic change between a maximum and a minimum value close to the
semimajor axis of the Earth for the tadpole and the horseshoe orbit.
In the middle graph the typical behaviour for a body in a horseshoe
orbit is shown which we can describe at the best in a rotating
coordinate system (comp. Fig.~\ref{schema}): close to the turning
point of the orbit the Trojan finds itself in the vicinity of the
Earth, the semimajor axes is larger than the one of the Earth and
then -- when the libation angle changes from about $10^{\circ}$ to
$350^{\circ}$ it reaches values smaller than $a_{Earth}$; now the
Trojans is on the other side of the Earth. In Fig.~\ref{3orbs} in
the lowest graphs an unstable orbit is shown which -- after leaving
to be in a horseshoe orbit -- is chaotic  with librations between
$0^{\circ}$ to $360^{\circ}$.

\subsection{Secular resonances}

In this part we present our preliminary results about the secular
resonances to understand the structure of the stability
diagram (Fig.~\ref{all-60}). A secular resonance happens when the
precession rate of an object ($\dot\varpi$ or $\dot\Omega$) equals
to one of the eigenfrequencies of the system \cite{mur99}. In a
secular resonance, the eccentricity or the inclination shows
long-term oscillations. The eigenfrequencies of the system can be
calculated through a linear analysis \cite{bre74} or can be
numerically determined \cite{lar90}.

\begin{figure}[h]
\centerline{
\includegraphics[width=8.5cm,angle=0]{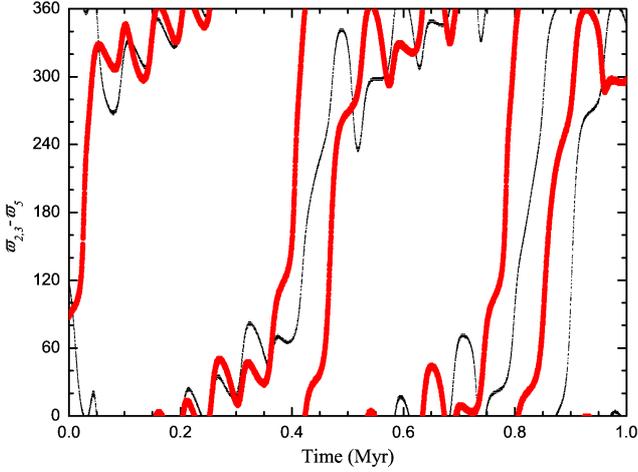}}
\caption{The differences between perihelion longitudes of inner
planets and of Jupiter. The abscissa is time and the ordinate is
$\varpi_k - \varpi_5$, with $k=2$ for Venus (thin black curve) and
$k=3$ for the Earth (thick red curve). }
 \label{force}
\end{figure}

The reality is more complicated than the model adapted in the linear
analysis. The secular perturbations among the planets modify the
precession rates. Due to their large masses the ones
of the big planets are almost constant; on the contrary, the precession rates
of the inner planets are strongly influenced
particularily by Jupiter and Saturn. In Fig.~\ref{force}, we show
the differences between Jupiter's perihelion longitude ($\varpi_5$,
as usually denoted) and the ones of Venus ($\varpi_2$) and the Earth
($\varpi_3$) in our simulations. Obviously, during the periods $\sim
0.05 - 0.35$~Myr and $\sim 0.45 - 0.75$~Myr, $\varpi_2$ and
$\varpi_3$ have nearly the same precession rate as Jupiter.
This makes the secular evolution of asteroids in this planetary
system complicated, as we will see below.

\begin{figure}[h]
\centerline{
\includegraphics[width=8.5cm,angle=0]{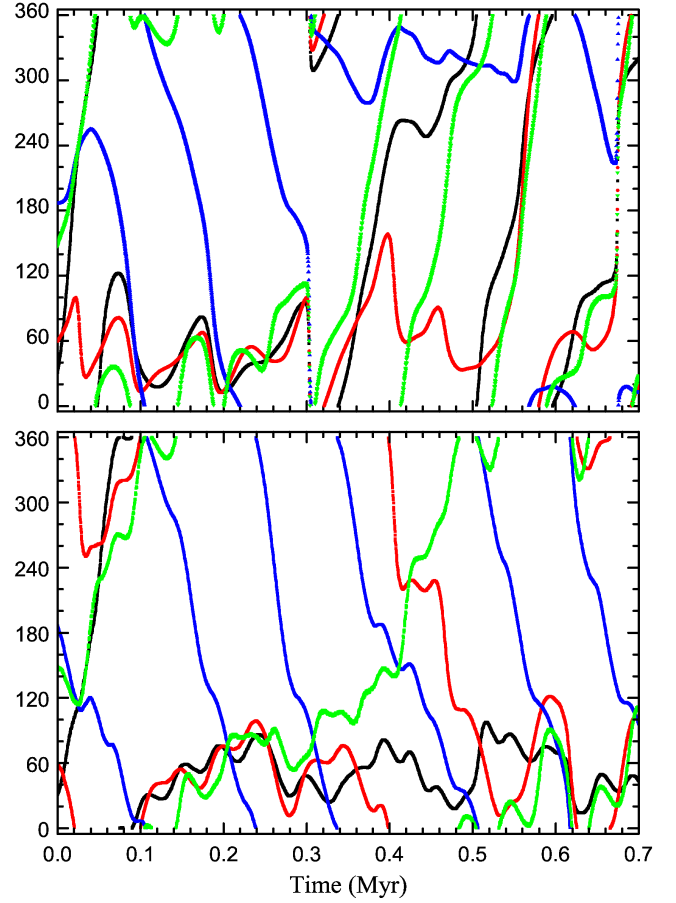}}
\caption{The evolution of apsidal differences between the Trojans
and the planets: $\Delta\varpi_2=\varpi-\varpi_2$
(black), $\Delta\varpi_3=\varpi-\varpi_3$ (red), $\Delta\varpi_4=\varpi-\varpi_4$
(blue) and $\Delta\varpi_5=\varpi-\varpi_5$ (green). The upper panel is for
$i=10^\circ$ and the lower panel is for $i=35^\circ$. }
 \label{resang1035}
\end{figure}

\begin{figure}[h]
\centerline{
\includegraphics[width=8.5cm,angle=0]{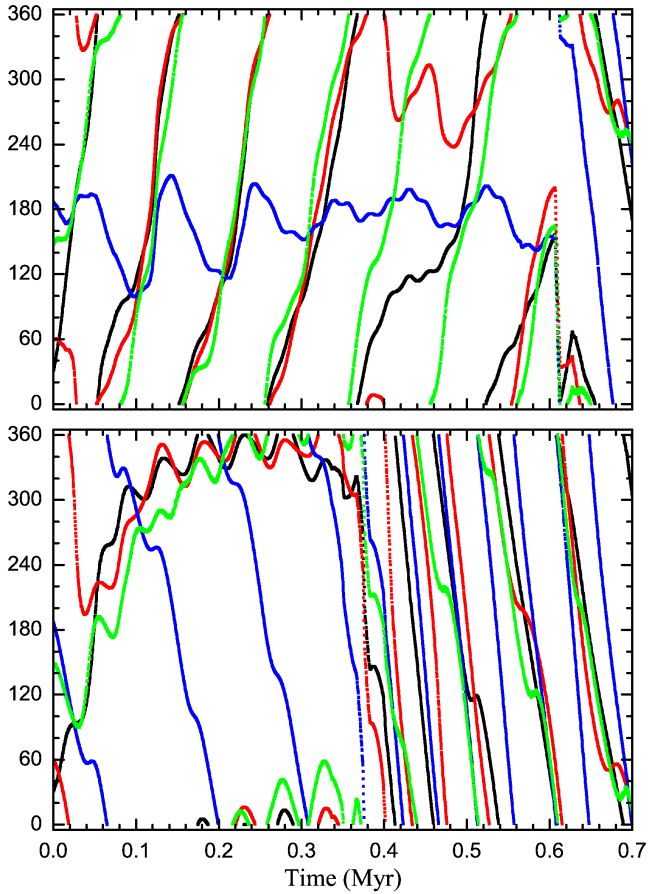}}
\caption{The evolution of apsidal differences between the Trojans
and the planets: $\Delta\varpi_2$ (black), $\Delta\varpi_3$ (red),
$\Delta\varpi_4$ (blue) and $\Delta\varpi_5$ (green). The upper
panel is for $i=22^\circ$ and the lower panel is for $i=42^\circ$.}
 \label{resang2242}
\end{figure}

\begin{figure}[h]
\centerline{
\includegraphics[width=8.5cm,angle=0]{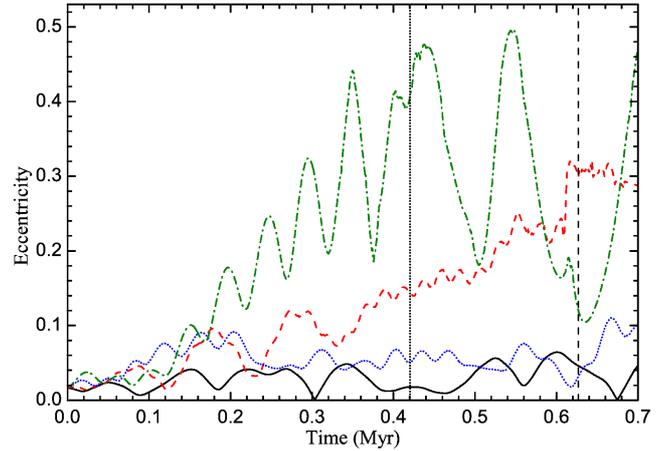}}
\caption{The eccentricity evolution for four orbits. The solid curves represent
the orbits with initial inclination $i=10^\circ$ (solid black), $22^\circ$
(dash red), $35^\circ$ (dot blue) and $42^\circ$ (dash-dot green),
respectively. The vertical dash and dot lines mark the moments when the orbits
escape from the 1:1 MMR for $i=22^\circ$ and $i=42^\circ$, respectively. }
 \label{eccvar}
\end{figure}

At high inclinations the Kozai resonance causes large coupled variations of
the eccentricity and the inclination \cite{Bra04}, which finally results in
instability. At low and moderate inclinations the secular
resonances $\nu_3$ and $\nu_4$ were found to affect the Earth
Trojans\footnote{A Trojan precesses at the rate of the
eigenfrequency $g_3$ (or $g_4$), which is the precession rate of the
Earth (or Mars), is in the $\nu_3$ (or $\nu_4$) secular resonance.}
\cite{Mor01, Bra02}. We notice there are
some disagreements in the locations of these secular resonances
between different papers. To depict a detail resonance map we would like to
leave it to a separate paper. Here we only illustrate the evolution of
four typical orbits to show the effects of secular resonances that
contribute to form the structures in the stability diagrams. In
Fig.~\ref{all-60} these are the four orbits with semi-major axis $a=0.9995$~AU and
inclination $i=10^\circ, 22^\circ, 35^\circ, 42^\circ$;
the evolutions of the resonant angles are presented in Figs.~\ref{resang1035} and
~\ref{resang2242}, respectively. The signal of the eccentricities can be
found in Fig.~\ref{eccvar}. Obviously, the orbits with $i=10^\circ$ and
$i=35^\circ$ are in the stable windows, while the orbits with
$i=22^\circ$ and $i=42^\circ$ are in the unstable gaps in
Fig.~\ref{all-60}.

Our preliminary calculations show that the most important secular
resonances affecting the Earth Trojans are $\nu_2, \nu_3, \nu_4$ and
$\nu_5$. According to the linear theory of secular perturbation, the
averaged temporal variation of the eccentricity of a celestial body
perturbed by a perturbing planet can be roughly approximated by

\begin{equation}
\label{eccres}
<\dot{e}> = C\sin(\Delta\varpi_i) = C\sin(\varpi - \varpi_i) \ ,
\end{equation}

\noindent where $C$ is a negative constant determined by the mass of
the planet and the Laplace coefficients \cite{mur99,lij06}, $\varpi$
and $\varpi_i$ are the perihelion longitudes of the celestial body
and the planet, and $\Delta\varpi_i = \varpi - \varpi_i$ with $i=1,...,5$.
If $0^\circ < \Delta\varpi_i < 180^\circ$
then $<\dot{e}>$ is less zero and if $180^\circ < \Delta\varpi_i < 360^\circ$
then $<\dot{e}>$ is greater zero.

For the orbit with $i=10^\circ$, from $t=0.05$~Myr to $0.3$~Myr the
angles $\Delta\varpi_2$, $\Delta\varpi_3$ and $\Delta\varpi_5$
librate, all around values $\sim 40^\circ$ (see Fig.~\ref{resang1035},
upper panel). Such libration affects
(but not excites, according
to Eq.~\ref{eccres}) the eccentricity of the Trojan. The overlap
of these secular resonances introduces chaos
to the motion, which can be seen in the irregular eccentricity
evolution (black in Fig.~\ref{eccvar}). After $t=0.3$~Myr the $\nu_2$ and
$\nu_5$ vanishes as the corresponding resonant angles circulate, but
$\Delta\varpi_4$ begins to librate around a value larger than
$180^\circ$, i.e. $\nu_4$ resonance appears. Although this $\nu_4$
resonance may increase the eccentricity since $\Delta\varpi_4 >
180^\circ$, this eccentricity pumping effect is offset by the $\nu_3$ resonance, which
is continuously present with the $\Delta\varpi_3 < 180^\circ$.

Similar evolution happens for $i=35^\circ$ (see Fig.~\ref{resang1035},
lower panel). In this case, before
$t=0.1$~Myr however, $\nu_3$ and $\nu_5$ librate around high values,
resulting in eccentricity excitation in this period (blue in Fig.~\ref{eccvar}).
The $\nu_4$ resonance does not happen at all, while $\nu_3$ and $\nu_5$ are
absent for about $0.1$~Myrs from $0.4$ to $0.5$~Myr. Most of time,
the resonant angles are smaller than $180^\circ$ and the
eccentricity stays low.

The situation is quite different for $i=22^\circ$ (see Fig.~\ref{resang2242},
upper panel). The $\nu_4$
resonance exists with the resonant angle around $180^\circ$ all the
way before the Trojan is expelled from the 1:1 MMR at $t \sim
0.63$~Myr. No other secular resonances are present until $t \sim
0.35$~Myr and the Trojan's eccentricity varies in correspondence with
the circulation of the main secular angles $\Delta\varpi_k$
($k=2,3,5$). The critical eccentricity excitation happens when the
$\nu_3$ resonance sets in at $t\sim 0.35$~Myr. In this secular
resonance the eccentricity reaches 0.25 at $t=0.55$~Myr (red in
Fig.~\ref{eccvar}), where the Trojan may approach to Venus closely.
And this destabilizes the Trojan orbit.

For $i=42^\circ$ (see Fig.~\ref{resang2242},
lower panel) the resonant angles of $\nu_2, \nu_3$ and $\nu_5$  librate
around high values, resulting in a quick excitation of the Trojan's
eccentricity to $0.45$ (red in
Fig.~\ref{eccvar}). Such a high eccentricity makes its
orbit to cross the one of Venus, the one of Mars and finally the
Trojan becomes unstable.

Based on these typical orbital evolutions we argue
that the secular resonances $\nu_2, \nu_3, \nu_4, \nu_5$ are all
involved in determining the Trojans' stability.
Among them the $\nu_3$ and $\nu_5$ play more important roles than
the others.

\section{Conclusions}

We investigated in full detail the stability of the recently found
Earth Trojan asteroid 2010 $\text{TK}_7$ in different dynamical
models. We derived a symplectic mapping based on the spatial
elliptic restricted three--body problem to see the location of the
mean orbit in the context of the phase space structure. We tested
the validity of the simplified model with numerical integrations and
found good agreement of the model on short time scales. We included
the influence of the additional planets as well as the Moon to see
their influences on the mean orbit of 2010 $\text{TK}_7$. Next, we
performed a detailed numerically study to propagate the orbit of the
Trojan Earth asteroid both, forward and backwards in time together
with 100 clone orbits to incorporate the possible errors of the
orbital parameters induced by the observations. From the detailed
study we are able  to state the probability of capture and escape of
the Earth Trojan together with the estimation of the time the
asteroid will stay close to the Lagrangian point $L_4$ of the Earth.
To this end we investigated in full detail not only the regime of
parameters and initial conditions close to the actual found asteroid
but rather on a grid of initial conditions and in the parameter
space $(a,i)$ using thousands of simulations, based on fictitious
Earth-Trojans. With this we were able to define the region of
stability and instability on both short and long time scales.

The main results of the present study can be summarized as follows:
we can confirm the result of \cite{Con11}, the asteroid 2010
$\text{TK}_7$ lies in the tadpole regime of the Sun-Earth system.
The orbital parameters indicate that most probably the asteroid
became an $L_4$ Trojan some 1,800 years ago and it will jump to the
$L_5$ neighbourhood or a horseshoe orbit in about 15,000 years in
future. Before it moved to the current tadpole orbit this asteroid
was captured into the 1:1 MMR around 30,000 years ago, and it may
stay in the resonance for another 200,000 years. The total life-time
of the asteroid (being in the 1:1 MMR) is less than $0.25$ Myr. As a
Near-Earth-Asteroid, the closest approach of this asteroid to the
Earth will be larger than $70$ times the Earth-Moon distance when it
is on a Trojan-like orbit. On short time scales it is possible to
predict the orbit in the spatial elliptic restricted three-body
problem (Sun-Earth-Asteroid) on intermediate time scales the
influence of the Moon has to be taken into account. In addition, on
longer time scales the influence of the other planets of our Solar
system cannot be neglected, since the asteroid is on a chaotic orbit
with close encounters to the unstable equilibrium $L_3$. We expect
the discovery of further interesting objects in the vicinity of the
$L_4$ or $L_5$ equilibria of the Sun-Earth system. The long term
integrations show: for low inclinations the stability region extends
in terms of semi-major axis up to $\pm.01$~AU. The size of it
decreases with increasing inclination up to a threshold at about
$i=20^{\circ}$. Another stability window opens up between
$28^{\circ}$ to $40^{\circ}$ and disappears soon after. The
preliminary results of our frequency analysis indicate that the
$\nu_2, \nu_3, \nu_4$ and $\nu_5$ secular resonances are deeply
involved in the motion of the Earth Trojans, and we show this by
several examples.

From a qualitative point of view our simulations indicate that 2010
$\text{TK}_7$ is situated inside an unstable region where all
sorts of orbits like horseshoes, tadpoles and jumping ones are
possible. We show -- this is well confirmed -- the Earth Trojan
moves on a temporary captured orbit. On the other hand it is
unclear why there are no other Trojans of the Earth found up to
now because there exist large stable regions for small
inclinations, but also for larger inclinations Trojans may exist
for very long time scales. \noindent

\acknowledgements{ZLY thanks the financial support by the National Natural Science Foundation of China
(No. 10833001, 11073012, 11078001).}

\end{document}